This is a pre-publication draft of a publication sponsored by the [Software Preservation Network's Law and Policy Working Group](#). We anticipate publishing a revised version in late 2021. Please send feedback to Ana Enriquez at aee32@psu.edu.

# Section 108 and Software Collections: A User's Guide

by Ana Enriquez

## Table of Contents

## Acknowledgments

Thank you to Brandon Butler, Kyle Courtney, Seth Erickson, Paula Jabloner, Raven Lanier, Alyssa Loera, Phil Salvador, Graeme Slaght, Nathan Tallman, and Yuanxiao Xu for helpful feedback.

Descriptions of the provisions of Section 108 in this guide are based on [Copyright Basics](), a guide from the University of Michigan Library, which is licensed under the Creative Commons Attribution 4.0 International License.

This user's guide explains Section 108 of the U.S. Copyright Act in the context of software collections. Linked terms lead to the glossary at the end of the document.

## Section 108 and Fair Use

Like fair use, Section 108 of the U.S. Copyright Act[1] provides important rights for libraries and archives engaged in software preservation. Unlike fair use, which is broad and flexible, Section 108 applies only to specific uses by qualified libraries and archives. Section 108 provides a helpful baseline, which is then augmented by fair use. In the places where it applies, Section 108 provides helpful certainty.

Section 108 and fair use operate together. The text of Section 108 and the case *Authors Guild v. HathiTrust*[2] tell us that libraries and archives can rely on both provisions; they need not choose between them. When a library or archives uses material in a way that's similar to what Section 108 permits, that use is more likely to be a fair use (even if it's not permitted by 108). For more information, see Jonathan Band's 2012 article, "The Impact of Substantial Compliance with Copyright Exceptions on Fair Use."[3]

For general information on fair use and software preservation, see the Code of Best Practices in Fair Use for Software Preservation.[4]

## Eligibility and Limitations: 108(a) and (g)

Section 108(a) describes which libraries and archives can rely on Section 108. A library or archives and its employees acting within the scope of employment qualify for 108 if:

1) the reproduction or distribution is not for direct or indirect commercial advantage;
2) the institution's collections are either
   a) open to the public or
   b) available to unaffiliated persons doing research in a specialized field; and
3) the reproduction or distribution includes the work's copyright notice (or, in the absence of a copyright notice, a substitute statement such as "This work may be protected by copyright.").

Many institutions engaged in software preservation qualify for these rights. While museums are not included, libraries and archives associated with museums can qualify.

One example of an eligible institution would be the Stanford University Library, which holds the Stephen M. Cabrinety Collection in the History of Microcomputing, among many other software holdings. Stanford's collection is open to the public and its preservation activities are not conducted for commercial advantage (direct or indirect). So long as the copies it makes include a copyright notice or substitute statement, it is eligible under 108(a).

---

[1] 17 U.S.C. § 108.

[2] *Authors Guild v. HathiTrust*, 755 F.3d 87, 94 fn. 4 (2d Cir. 2014).

[3] Jonathan Band, "The Impact of Substantial Compliance with Copyright Exceptions on Fair Use," *Journal of the Copyright Society of the USA* 59, no. 3 (Spring 2012): 453-476.

[4] Center for Media and Social Impact, *Code of Best Practices in Fair Use for Software Preservation* (rev. 2019).

By contrast, an example of an ineligible institution might be a corporate archive that is accessible only to employees, such as the software archives at Apple Computer. Unless it is open to researchers unaffiliated with Apple, the archive does not meet the first requirement of 108(a). (Of course, if you own all the copyrights in the collection, as Apple does in this scenario, Section 108 might not be an issue for you!)

Section 108(g) states that the rights in Section 108 do not apply if the library or archives, or its employee, "is aware or has substantial reason to believe that it is engaging in the related or concerted reproduction or distribution of multiple copies or phonorecords of the same material, whether made on one occasion or over a period of time, and whether intended for aggregate use by one or more individuals or for separate use by the individual members of a group." For example, if a library employee learns that students in a particular class were all making requests for the same material, Section 108 would not apply to those requests. However, the library could still analyze the requests under fair use and might be able to meet the class's needs that way.

108(g) also clarifies that institutions can make unrelated copies of the same material on multiple occasions. For example, under 108(c) a library could make up to three copies of a software program for the purpose of replacement of a damaged copy. It could later make a copy of the same program for a patron under 108(e).

## Copying Unpublished Works for Preservation and Security: 108(b)

A qualified library or archives is entitled to make up to three copies of an unpublished work currently in its collection "solely for purposes of preservation and security or for deposit for research use in another" qualified library or archives, so long as any copy "that is reproduced in digital format is not otherwise distributed in that format and is not made available to the public in that format outside the premises of the library or archives."

For example, an archives that holds an unpublished typescript software manual could create a physical copy of that manual at the request of another archives. The second archives could acquire that copy as part of its collection. (In the language of the statute, this is "deposit for research use.")

In another example, an archives that holds an unpublished piece of research software on floppy disk could copy that software for "purposes of preservation and security" at the original institution. Because the copy would be digital, 108(b) would only permit the archives to make the copy available to the public on the premises of the archives. This would permit reading room access to the copy, for example.

## Copying Published Works to Create Replacement Copies: 108(c)

A qualified library or archives is entitled to make up to three copies of a published work "solely for the purpose of replacement of a copy or phonorecord that is damaged, deteriorating, lost, or stolen, or if the existing format in which the work is stored has become obsolete," so long as an unused replacement copy of the work cannot be obtained at a fair price and any copy "that is reproduced in digital format is not made available to the public in that format outside the premises of the library or archives in lawful possession of such copy."[5]

For example, suppose a library holds a copy of the game Oregon Trail 3 on CD, and the particular copy is deteriorating due to disc rot. Oregon Trail 3 is a published work. So long as the library determines that an unused replacement copy cannot be obtained at a fair price, it can make up to three copies as replacement copies. For example, the library could store two copies in back-up storage and provide one copy as an access copy. Because the replacement copies would be in digital formats, 108(c) would not permit them to be made available to the public outside the library premises, but they could be used on site.

## Copying at the Request of Users: 108(d) and (e)

Upon request from one of its users or the user of another qualified library or archives, a qualified library or archives is entitled to make a single copy of

1) per 108(d), "no more than one article or other contribution to a copyrighted collection or periodical issue;"
2) "a small part of any other copyrighted work;" or
3) per 108(e), an "entire work, or . . . a substantial part of it . . . if the library or archives has first determined, on the basis of a reasonable investigation, that a copy or phonorecord of the copyrighted work cannot be obtained at a fair price,"

So long as:
1) that copy becomes the property of the requesting user;
2) the institution has no notice that the copy or phonorecord would be used for any purpose other than private study, scholarship, or research;
3) the institution displays a copyright warning in accordance with regulations from the US Copyright Office (see Implementing Section 108 at Your Institution for details); and
4) per 108(i), the work copied is not a musical work, a pictorial, graphic or sculptural work, or a motion picture or other audiovisual work other than an audiovisual work dealing with news. (However, "pictorial or graphic works published as illustrations, diagrams, or similar adjuncts to works" can be copied.)

In addition, per 108(g), the qualified library or archives may not "engage[] in the systematic reproduction or distribution of single or multiple copies or phonorecords of material described in subsection (d)," that is, materials for which a reasonable investigation has not been conducted or where that investigation found a copy available at a fair price. However, this clause does not prevent qualified libraries or archives from participating in interlibrary loan, so long as their participation does not substitute for a subscription or purchase of the loaned works.

---

[5] For an overview of 108(c) for video collections, see Howard Besser, Melissa A. Brown, Robert Clarida, Walter Forsberg, Mark Righter, and Michael Stoller, "Video at Risk: Strategies for Preserving Commercial Video Collections in Libraries," December 2012. Although the video-specific information is unlikely to apply directly to most software collections, this is an excellent resource for understanding 108(c) and its terms in practice.

For example, suppose a patron requests a copy of several functions from a Python script held by an archives. This would qualify as "a small part of any other copyrighted work." If the archives has followed the steps in [Implementing Section 108 at Your Institution](#) and has no notice that the copy would be used for a purpose other than private study, scholarship, or research, it can make a copy and provide it to the patron.

More often, software researchers are likely to request entire scripts or programs. If a patron requests a copy of the entire Python script, the archives will first need to conduct a [reasonable investigation](#). If that investigation determines that a copy of the copyrighted work cannot be obtained at a [fair price](#), the archives can copy the entire script for the patron.

Due to the limitation in 108(i), 108(d) and (e) apply to only certain types of copyrighted works. Many pieces of software, including the Python script imagined here, are considered "literary works" under U.S. copyright law. Literary works are eligible for reproduction under 108(d) and (e). Some pieces of software, such as video games, may be audiovisual works, which are not eligible for reproduction under 108(d) and (e). No court has ruled on this question in the context of Section 108. For more information, see the [glossary entry on motion pictures and other audiovisual works](#).

## Displaying Notices on Unsupervised Reproducing Equipment: 108(f)(1)

Under U.S. copyright law, a person or entity (such as a library or archives) is only liable for someone else's (such as a patron's) infringement under specific circumstances. Those circumstances are exceedingly rare.[6] Under 108(f)(1), qualified libraries and archives can further reduce their liability by displaying a notice on unsupervised reproducing equipment that states that "the making of a copy may be subject to copyright law." Many libraries and archives choose to share information about users' rights, such as fair use, at the same time.

Unsupervised reproducing equipment common in libraries and archives that collect software includes scanners and photocopiers (for copying printed materials that accompany software), computer terminals, and peripherals such as optical drives. Only equipment that is available for unsupervised use by patrons should bear these notices. There is no benefit to displaying these notices on equipment that is used only by staff.

---

[6] ["Should Libraries Fret Over Mischievous Users?," ARL Policy Notes Blog (March 16, 2012).](#)

# Using Works During the Last Twenty Years of Their Copyright Terms: 108(h)

During the last 20 years of a work's copyright term, a qualified library or archives in entitled to "reproduce, distribute, display, or perform in facsimile or digital form a copy or phonorecord of such work, or portions thereof, for purposes of preservation, scholarship, or research, if such library or archives has first determined, on the basis of a reasonable investigation" that

1) The work is not subject to normal commercial exploitation;
2) A copy of the work cannot be obtained at a reasonable price; and
3) The copyright holder has not provided notice of the work being subject to commercial exploitation or available at a reasonable price.[7]

Under U.S. law, copyright term for 20th-century materials is lengthy and complicated. "Copyright Term and the Public Domain in the United States," a chart created by Peter Hirtle and maintained by the Cornell University Library Copyright Information Center, is a useful resource for understanding U.S. copyright duration.[8] While some software is in the public domain, very little software is currently in the last twenty years of its copyright term.

A work is in the last twenty years of its copyright term as of 2021 if its term expires on or before January 1, 2041. For published software, this would mean a publication date of 1945, which is very early for software. Early published software from the late 1940s and early 1950s will enter the last twenty years of their copyright terms in the coming decade.

In contrast, unpublished works are under copyright for 70 years after the death of the last surviving author. As of 2021, unpublished software by authors who died in 1970 or earlier is already in the last twenty years of its copyright term. Unpublished software by authors who died in 1950 or earlier is in the public domain. For unpublished works of corporate authorship (works made for hire), the term is 120 years from date of creation. Unpublished work-for-hire software from the mid 20th century will not be in the last twenty years of its term until the mid 21st century.

# Implementing Section 108 at Your Institution

## General steps for implementation of Section 108 at your institution

1. Confirm that your institution meets the eligibility requirements from 108(a), summarized above. Stay within the parameters for the particular subsection you're using.

---

[7] Per 37 C.F.R. § 201.39, rightsholders can file a notice with the U.S. Copyright Office indicating that a work is subject to normal commercial exploitation or available at a reasonable price. As of the September 2017 publication United States Copyright Office, "Section 108 of Title 17: A Discussion Document of the Register of Copyrights," no such notices had ever been filed.

[8] Cornell University Library Copyright Information Center, "Copyright Term and the Public Domain in the United States" (rev. 2021).

2. When reproducing or distributing a work under any part of Section 108, retain the work's copyright notice. In the absence of a copyright notice, add a substitute statement such as "This work may be protected by copyright."
3. Do not rely on Section 108 if:
    a. You become "aware or [have] substantial reason to believe that [your institution] is engaging in the related or concerted reproduction or distribution of multiple copies or phonorecords of the same material." For example, 108 should not be used to copy the same work repeatedly for students in an academic course. (Note that fair use can still apply in this situation.)

### Additional steps for uses under Section 108(d) and (e)

Additional steps are necessary when taking advantage of 108(d) and (e). Title 37, section 201.14 of the Code of Federal Regulations[9] supplements 108 to provide detailed requirements.

1. Use the following text, verbatim, for the Display Notice and Order Notice described below. In addition to the required text, you may provide other information about copyright.

    > Notice Warning Concerning Copyright Restrictions
    >
    > The copyright law of the United States (title 17, United States Code) governs the making of photocopies or other reproductions of copyrighted material.
    >
    > Under certain conditions specified in the law, libraries and archives are authorized to furnish a photocopy or other reproduction. One of these specific conditions is that the photocopy or reproduction is not to be "used for any purpose other than private study, scholarship, or research." If a user makes a request for, or later uses, a photocopy or reproduction for purposes in excess of "fair use," that user may be liable for copyright infringement.
    >
    > This institution reserves the right to refuse to accept a copying order if, in its judgment, fulfillment of the order would involve violation of copyright law.

2. "At the place where orders for copies or phonorecords are accepted," such as a special collections desk or interlibrary loan desk, display a Display Notice of Copyright, using the "Notice Warning Concerning Copyright Restrictions" text included above. This notice must "be printed on heavy paper or other durable material in type at least 18 points in size, and . . . be displayed prominently, in such manner and location as to be clearly visible, legible, and comprehensible to a casual observer within the immediate vicinity of the place where orders are accepted." If your library also accepts orders online, we recommend displaying a similar notice on the website where patrons place orders.

---

[9] 37 C.F.R. § 201.14.

3. On forms (print and/or digital) used by patrons ordering copies, include an Order Notice of Copyright, using the "Notice Warning Concerning Copyright Restrictions" text included above. This notice must "be printed within a box located prominently on the order form itself, either on the front side of the form or immediately adjacent to the space calling for the name or signature of the person using the form. The notice shall be printed in type size no smaller than that used predominantly throughout the form, and in no case shall the type size be smaller than eight points. The notice shall be printed in such manner as to be clearly legible, comprehensible, and readily apparent to a casual reader of the form."
4. Do not rely on Section 108(d) or (e) if you have "notice that the copy or phonorecord would be used for any purpose other than private study, scholarship, or research." Note that this provision does not prevent users who receive the copies from making other noninfringing uses of the work. However, if a patron indicates to a library or archives employee that they will use the copy for a purpose other than private study, scholarship, or research, the library or archives should not rely on 108(d) or (e) to make the copy. Fair use remains available, if it applies to the use in question.

## Glossary

**Commercial advantage**

Reproduction and distribution that relies on Section 108 must be done "without any purpose of direct or indirect commercial advantage." Libraries and archives at non-profit institutions always meet this requirement. Section 108 can also apply to reproduction and distribution by libraries and archives that are part of commercial entities (e.g., law firm libraries or archives at for-profit publishing companies, architectural firms, or media outlets), so long as the reproduction or distribution does not provide direct or indirect commercial advantage and they meet the other eligibility requirements.[10] For example, if it meets the other eligibility

---

[10] In House Report No. 94-1476, the Committee on the Judiciary of the House of Representatives provided additional information about the intention behind various provisions of the Copyright Act, including Section 108. Courts interpreting federal statutes consider, and try to bring about, Congress's intent. Beginning on page 74, the House Report provides guidance on the question of "direct or indirect commercial advantage." The report connects the prohibition on commercial advantage to the prohibition on "systematic" reproduction from 108(g):

> Isolated, spontaneous making of single photocopies by a library in a for-profit organization, without any systematic effort to substitute photocopying for subscriptions or purchases, would be covered by section 108, even though the copies are furnished to the employees of the organization for use in their work. Similarly, for-profit libraries could participate in interlibrary arrangements for exchange of photocopies, as long as the production or distribution was not "systematic." These activities, by themselves, would ordinarily not be considered "for direct or indirect commercial advantages," since the "advantage" referred to in this clause must attach to the immediate commercial motivation behind the reproduction or distribution itself, rather than to the ultimate profit-making motivation behind the enterprise in which the library is located. On the other hand, section 108 would not excuse reproduction or distribution if there were a commercial motive behind the actual making or distributing of the copies, if multiple copies were made or distributed, or if the photocopying activities were "systematic" in the sense that their aim was to substitute for subscriptions or purchases.

requirements in 108(a), the archives of a for-profit scientific research firm could likely rely on Section 108(b) to create a copy of unpublished research software from its collection to be added to the collection of another qualified library or archives.

## Copies and phonorecords

Per Section 101 of the Copyright Act, "'Copies' are material objects, other than phonorecords, in which a work is fixed by any method now known or later developed, and from which the work can be perceived, reproduced, or otherwise communicated, either directly or with the aid of a machine or device. The term 'copies' includes the material object, other than a phonorecord, in which the work is first fixed." The definition of "phonorecord" is very similar, except that "sounds" are the things fixed: "'Phonorecords' are material objects in which sounds, other than those accompanying a motion picture or other audiovisual work, are fixed by any method now known or later developed, and from which the sounds can be perceived, reproduced, or otherwise communicated, either directly or with the aid of a machine or device. The term 'phonorecords' includes the material object in which the sounds are first fixed."

## Fair price and reasonable price

"Fair price" appears twice, in 108(c) and (e). "Reasonable price" appears in 108(h). These phrases likely have similar meanings. No court has interpreted this language. Some experts have suggested that if a work could only be purchased in a larger bundle of works, that price would not be fair. Another common belief is that the availability of a first edition, autographed copy, or an otherwise special copy at a premium would not constitute the work being available at a fair price.

## Motion picture or other audiovisual work

Per Section 101 of the Copyright Act, "'Audiovisual works' are works that consist of a series of related images which are intrinsically intended to be shown by the use of machines, or devices such as projectors, viewers, or electronic equipment, together with accompanying sounds, if any, regardless of the nature of the material objects, such as films or tapes, in which the works are embodied. . . . 'Motion pictures' are audiovisual works consisting of a series of related images which, when shown in succession, impart an impression of motion, together with accompanying sounds, if any." No court has interpreted this provision in the context of 108. The practices of the U.S. Copyright Office in registering works are a source of potential guidance here. The Office considers "arcade games and videogames" to be audiovisual works. "Applications designed for mobile phones and tablets" may also be audiovisual works.[11]

## Musical work

The Copyright Act does not define "musical work," but there is an administrative definition from the U.S. Copyright Office: "For purposes of copyright registration, musical works (which are also known as musical compositions) are original works of authorship consisting of music and any accompanying words. Music is a succession of pitches or rhythms, or both, usually in some definite pattern."[12] Musical works are the musical composition, rather than the "sound recording," which is a recording of a particular performance of a piece of music. For example, software containing sheet music for copyrighted songs (e.g., for practicing transposition) includes musical works and thus would be excluded from copying under 108(d) and (e). Software containing a sound recording of The Chicks' cover of "Landslide" would also be excluded, because the sound recording embodies the musical work. In contrast, software containing a recording of a Bach cantata would likely not be excluded, because the Bach cantata is in the public domain, so you would not need to rely on 108 for copying the musical work (though you would for the rest of the software).

---

[11] *The Compendium of U.S. Copyright Office Practices*, Chapter 800, §§ 807.1 & 807.7(C).

## Obsolete

Section 108(c) defines obsolete: "For purposes of this subsection, a format shall be considered obsolete if the machine or device necessary to render perceptible a work stored in that format is no longer manufactured or is no longer reasonably available in the commercial marketplace." With software, the carrier/storage media (e.g., floppy disk) could be obsolete, as could the file format in which the software is stored. For example, a file format that can only be rendered in Mac OS 9 would be considered obsolete because machines necessary to render it perceptible (i.e., Macs that can run OS 9 or run in Classic mode) are no longer manufactured and no longer reasonably available in the commercial marketplace. This would be the case even if the program is stored in a non-obsolete storage medium such as a CD.

Because emulators do not render works perceptible to the same extent that original equipment did, the availability of an emulator will probably not prevent a format from being considered obsolete. However, the manufacture or commercial availability of reproduction hardware that fully renders perceptible works in a particular format likely prevents that format from being considered obsolete. Works in such formats may nonetheless be eligible for copying under 108(c) due to being "damaged, deteriorating, lost, or stolen."

## Pictorial, graphic, or sculptural work

Per Section 101 of the Copyright Act, "'Pictorial, graphic, and sculptural works' include two-dimensional and three-dimensional works of fine, graphic, and applied art, photographs, prints and art reproductions, maps, globes, charts, diagrams, models, and technical drawings, including architectural plans." While Section 108(d) and (e) generally do not apply to these types of works, they do allow copying of "pictorial or graphic works published as illustrations, diagrams, or similar adjuncts to works" permitted to be copied under those sections. Many pictorial and graphic works included in software (e.g., icons in a word processing program) are included as these sorts of adjuncts.

---

[12] *Id.* at § 802.1.

## Publication

A work is "published" for the purposes of the Copyright Act if copies of the work have been distributed to the public with the authorization of the copyright holder. Much software is published, but some is not. For example, if a researcher provided a copy of their research software to their institution's library for preservation, this alone is unlikely to constitute publication. In contrast, if a researcher handed out CDs containing their software at a conference, that would constitute publication.

## Reasonable investigation

The Copyright Act does not define reasonable investigation. For commercially distributed software, it likely makes sense to check whether the original distributor is still selling copies or licenses to the software. For certain types of software, there may be popular third-party vendors to check as well. 108(c) requires a search for an unused replacement. For (e), which does not specify unused, it may also make sense to check eBay or other used software sales sites.